# Stress evolution in lithium-ion composite electrodes during electrochemical cycling and resulting internal pressures on the cell casing

Siva P.V. Nadimpalli,[a] Vijay A. Sethuraman,[b] Daniel P. Abraham,[c] Allan F. Bower,[b] Pradeep R. Guduru[b,*]

[a] *Department of Mechanical and Industrial Engineering, New Jersey Institute of Technology, Newark, NJ 07940, USA*

[b] *School of Engineering, Brown University, Providence, RI 02912, USA*

[c] *Chemical Sciences and Engineering Division, Argonne National Laboratory, Argonne, IL 60439, USA*

\* Corresponding author. Email: Pradeep_Guduru@Brown.edu; Tel: 401 863 3362

**Abstract**

Composite cathode coatings made of a high energy density layered oxide ($Li_{1.2}Ni_{0.15}Mn_{0.55}Co_{0.1}O_2$, theoretical capacity ~377 mAh-g$^{-1}$), polyvinylidene fluoride (PVdF) binder, and electron-conduction additives, were bonded to an elastic substrate. An electrochemical cell, built by pairing the cathode with a capacity-matched graphite anode, was electrochemically cycled and the real-time *average* stress evolution in the cathode coating was measured using a substrate-curvature technique. Features in the stress evolution profile showed correlations with phase changes in the oxide, thus yielding data complementary to *in situ* XRD studies on this material. The stress evolution showed a complex variation with lithium concentration suggesting that the volume changes associated with phase transformations in the oxide are not monotonically varying functions of lithium concentration. The peak tensile stress in the cathode during oxide delithiation was approximately 1.5 MPa and the peak compressive stress during oxide lithiation was about 6 MPa. Stress evolution in the anode coating was also measured separately using the same technique. The measured stresses are used to estimate the internal pressures that develop in a cylindrical lithium-ion cell with jelly-roll electrodes.

**Key words**

Layered oxide, electrochemical cycling, *in situ* stress measurement, jelly-roll electrodes, internal pressure, and cylindrical cell.

## 1. Introduction

Lithium-ion cells are the primary choice for portable energy storage devices because of their high energy densities. Presently, lithium-ion cells typically contain carbon-based materials, such as graphite, in the negative electrode (anode) and transition metal oxides in the positive electrode (cathode). State-of-the-art lithium-ion batteries have a specific capacity of ~ 150Wh-$kg^{-1}$; energy densities, two to five times higher, are needed to meet the performance and range requirements of hybrid and all-electric vehicles for transportation applications [1]. The energy densities can be increased by using anode and cathode materials with higher capacities and/or higher voltages. For example, anode materials Sn and Si can accommodate several lithium atoms per metal/metalloid unit yielding theoretical capacities of 960 ($Li_{4.25}Sn$) and 4009 mAh-$g^{-1}$ ($Li_{4.2}Si$), respectively. Cathode materials from the $x$$Li_2MnO_3$•($1$-$x$)$LiMO_2$ family of compounds have been reported to display capacities approaching 300 mAh-$g^{-1}$, significantly larger than the 140-170 mAh-$g^{-1}$ useable capacities exhibited by commercial materials.

Several recent investigations have focused on exploring the structure-property relationships of the $x$$Li_2MnO_3$•($1$-$x$)$LiMO_2$ compounds. Multiple experimental techniques including electron microscopy [2, 3], X-ray and neutron diffraction [4-6], X-ray absorption spectroscopy [7], and nuclear magnetic resonance [8, 9] have been used to characterize the structural changes in these oxides resulting from electrochemical cycling. In addition to structural changes, variation in lithium concentrations during electrochemical cycling can result in complex stress fields within (and between) the oxide secondary particles resulting in mechanical degradation [10], which can affect performance and life of the cathode coatings [11, 12]. The objective of this work is to measure the real-time thickness-averaged-stress evolution in composite cathode coatings during electrochemical cycling. This effort complements previous studies that examined the real-time stress evolution in anode materials and anode coatings during electrochemical cycling [13-15].

In this work we report on stress evolution measurements in composite cathode coatings containing $Li_{1.2}Ni_{0.15}Mn_{0.55}Co_{0.1}O_2$ ($0.5Li_2MnO_3$•$0.5LiMn_{0.375}Ni_{0.375}Co_{0.25}O_2$) as the active material using the experimental method developed by Sethuraman *et al.* [13]. The cathode was paired with a capacity-matched graphite-based anode to mimic the conditions in a practical cell. We note that several features in the stress evolution profiles are correlated with phase changes in the oxide during the lithiation/delithiation process determined by *in situ* X-ray diffraction (XRD) measurements [9, 16]. We also report the stress evolution in the graphite anode coating that was measured separately using the same technique. These stress evolution data are then used to estimate the internal pressures developed in a cylindrical lithium-ion cell containing spirally-wound electrodes during a charge and discharge cycle.

## 2. Experimental Methods

### 2.1 Sample Fabrication

The positive and negative electrode formulations are shown in Table 1. In addition to the active material $Li_{1.2}Ni_{0.15}Mn_{0.55}Co_{0.1}O_2$ (theoretical capacity of 314 mAh-$g^{-1}$ and 377 mAh-$g^{-1}$ corresponding to delithiation states of $Li_{0.2}Ni_{0.15}Mn_{0.55}Co_{0.1}O_2$ and $Ni_{0.15}Mn_{0.55}Co_{0.1}O_2$, respectively), the 35 μm thick positive electrode coating also contains conductive carbon black (SuperP) and graphite for improved electronic conductivity and PVdF for binding the materials together and for adhesion to a 15 μm thick Al current collector. Figs. 1 (a) and (b) show that the $Li_{1.2}Ni_{0.15}Mn_{0.55}Co_{0.1}O_2$ secondary particles comprise agglomerates of rod and disc shaped primary particles. Furthermore, Table 1 indicates that oxide loading in the electrode is 6.64 mg-cm$^2$ and electrode porosity is 37.1%. The negative electrode coating (see Fig. 1c) contains graphite, SuperP for improved electronic conductivity, and a PVdF binder for coating cohesion and for adhesion to a 10 μm thick Cu current collector.

The electrode samples were cut into circular discs of 50.8 mm diameter and epoxy bonded to the rough side of single-side-polished (111) Si wafers (nominal thickness 450 μm); a schematic and SEM image of the assembly are shown Figs. 2(a) and 2(b). The role of the Si wafer is to serve as an elastic substrate which undergoes curvature change in response to stress in the electrode bonded to it; as such, the substrate may be replaced by any elastic material that does not participate in the electrochemical reactions.

## 2.2 Electrochemical Experiments

The electrochemical cells were assembled and tested in an argon-atmosphere glove box ($O_2$, $H_2O$ < 1 ppm) to minimize the impact of moisture and oxygen. All cells contained a Celgard polymer separator and an electrolyte solution of 1 M lithium hexafluorophosphate ($LiPF_6$) in a 1:1:1 ratio of ethylene carbonate (EC): diethyl carbonate (DEC): dimethyl carbonate (DMC). Electrochemical cycling was conducted with a Solartron 1470 E Multistat. In our experiments the stresses developed in the cathode coating were measured in a full cell configuration, that is, with the graphite-based anode. The full cell was charged and discharged galvanostatically at a current density of 0.1 mA/cm$^2$ (~C/20 rate) between 2.0 and 4.7 V. The stresses developed in the anode coating were measured in a half cell configuration, that is, with a Li-metal counter electrode; the cell was galvanostatically cycled between 1.2 and 0.01 V vs. Li/Li$^+$ at a ~C/20 rate.

## 2.3 *In situ* Measurement of Stress in Electrode Coatings

A state of equi-biaxial stress is induced in a film (or coating) on a substrate when the latter constrains the in-plane volume change of the former during electrolyte wetting or electrochemical cycling. The stress evolution in our electrode coatings was measured by monitoring the substrate curvature with a multi-beam optical sensor (MOS) setup (k-Space Associates, Dexter, MI). As illustrated in Fig. 2(c) the electrodes are arranged such that the polished side of the silicon substrate is used for curvature measurements in MOS setup. The MOS system consists of a laser source that generates a collimated beam, two etalons arranged orthogonally to each other to generate a 2 X 2 array of parallel beams, and a CCD camera to capture the beam array reflected from the sample surface. The relative change in the distance between the laser spots on the CCD screen gives the sample curvature, $\kappa$, as

$$\kappa = \frac{\cos\phi}{2L}\left\{\frac{D_o - D}{D_o}\right\}, \quad\ldots\ldots\ldots\ldots\ldots\ldots\ldots \text{Eq. 1}$$

where $D$ is the distance between the laser spots and $D_o$ is its initial value, $\phi$ is the reflection angle of the beam, and $L$ is the optical path length from the sample to the CCD camera. The factor (*cos $\phi$*)/2L, known as mirror constant, is specific to a setup and is obtained by calibrating the system in Fig. 2 with a reference mirror of known curvature in the sample plane. The 2 X 2 array of reflected spots enables curvature measurement in two orthogonal directions.

The Stoney equation is widely used to determine stresses in thin films on substrates determined by the substrate curvature method [17, 18]. However, in the present study, the cathode and anode coatings cannot be considered as thin films because their thickness is not negligible compared to that of the substrate. In addition, the stiffness contribution from additional layers, such as the epoxy and (Al or Cu) current collector, needs to be considered when relating the average coating stresses to the measured substrate curvature. Sethuraman *et al.* [13] used a modified Stoney equation to account for the multiple layers shown in Fig. 2 (a), which is given in Equation 2.

$$\sigma = \frac{M_1 h_1^2 \kappa}{6 h_4 f(h_i, M_i)} \quad \text{................................ Eq.2}$$

where $\sigma$ is the average biaxial stress in the electrode coating, $h_i$ and $M_i$ are the coating thicknesses and bi-axial moduli, respectively; $f(h_i, M_i)$ is a function of the thickness and biaxial modulus of all the layers shown in Fig.2 (a) and is given in Ref. [13]. It should be noted that for the cathode, the coating thickness, $h_4$, changes slightly due to oxide phase changes associated with lithium insertion/extraction during electrochemical cycling. However, in the present study, we assume $h_4$ to be constant, which is reasonable given that the volume change in layered oxides is quite small [11, 19]. Similarly, we assume that the anode coating thickness is constant during the lithiation and delithiation processes; this assumption is consistent with our previous data obtained on graphite-based electrodes [13]. It should also be noted that the actual 3-D stress distribution in the coating would be highly non-homogeneous due to the complex microstructure and multiple constituents of the coating (Fig. 1); the stress $\sigma$ in Eq. 2 should be interpreted as the average in-plane normal stress that represents the thickness average of the actual 3D stress field, which is the quantity measured and discussed in this paper.

## 3  Results and discussion

### 3.1 Stress and electrochemical response of composite cathode and anode coatings

Fig. 3 shows stress evolution in the cathode coating on wetting by the electrolyte, i.e., prior to any electrochemical cycling. Solvent absorption by the coating induces this stress, which is measured through substrate curvature change immediately after introducing electrolyte into the electrochemical cell. The stress evolution is approximately linear at the beginning and reaches a steady-state value of about -1 MPa (Fig. 3) after ~ 60 min (note that compressive stresses are shown as negative values and tensile stresses are shown as positive values). The stress resulting from solvent absorption is a function of the composite electrode constitution, binder material, tortuosity, porosity, etc.; the type and nature of the active constituents (i.e., oxide or graphite) does not appear to significantly influence this stress. For instance, Sethuraman et al. [13] observed a similar compressive stress response, and measured a value of ~ -1.25 MPa, during electrolyte absorption in a PVdF-based graphite anode.

After soaking the electrodes in electrolyte for ~60 min, the full cell was galvanostatically cycled at a ~C/20 rate. Figs. 4(a) and 4(c) show stress evolution in the cathode coating and full cell voltage, respectively, during the first charge/discharge cycle as a function of time; Figs. 4(b) and 4(d) show the same information as a function of oxide capacity. Fig. 4(d) shows that the first cycle charge capacity is ~311 mAh-$g_{oxide}^{-1}$ (corresponds to a stoichiometry of $Li_{0.21}Ni_{0.15}Mn_{0.55}Co_{0.1}O_2$); electrochemical activation of the oxide particles is seen as a ~4.4V pseudo-plateau during this cycle. This activation is typically associated with structural changes and loss of oxygen from the oxide [20]. The first cycle discharge capacity is ~250 mAh-$g_{oxide}^{-1}$ (corresponds to a stoichiometry of $Li_{1.0}Ni_{0.15}Mn_{0.55}Co_{0.1}O_2$), which is comparable to that obtained in coin-cell experiments [20]. Figs. 4 (a) and (b) show that the stress evolution is not monotonic indicating complex changes in the electrode during the first charge/discharge cycle.

In electrode coatings bonded to substrates, volume contraction of particles leads to tensile stress and volume expansion leads to compressive stress. The stress during the first charge cycle is entirely tensile indicating volume contraction of oxide particles in the electrode. Fig. 4(b), however, shows that the stress increases initially (stage I) until the cell voltage is ~3.9V, then is relatively stable at ~1.2 MPa until the cell voltage is ~4.35V. On further delithation, at the ~4.4V pseudo-plateau, the stress decreases (Stage II) and reaches ~0.2 MPa (at ~220 mAh-$g_{oxide}^{-1}$). Further delithiation increases the electrode stress to ~1.5 MPa till the cell capacity reaches ~280 mAh-$g_{oxide}^{-1}$ (Stage III); then the stress decreases to almost zero at 4.7V and ~311 mAh-$g_{oxide}^{-1}$ capacity (Stage IV).

The complex behavior described above is partially related to lattice parameter changes in $Li_{1.2}Ni_{0.15}Mn_{0.55}Co_{0.1}O_2$ during the initial delithiation of a similar oxide as described by previous investigators [16, 21]. Their *in operando* X-ray diffraction measurements reveal an increase in *c*- and decrease in *a*- lattice parameters leading to a gradual non-monotonic decrease in unit cell volume during the first charge to 4.8V. The observed crystal structure changes do not correlate exactly with the measured stress, which indicates that the supposedly 'inactive' electrode components such as the carbons and the binder also contribute to stress evolution. For instance, intercalation of $PF_6^-$ anions at voltages >4.45V vs. $Li/Li^+$ leads to lattice expansion and structure disordering of graphite contained in the positive electrode [22]. In addition, the maximum stress, between 1.2 and 1.5 MPa, displayed by the positive electrode could also be due to the finite tensile strength of the PVdF binder.

During the first discharge cycle, the stress profile shows a brief initial increase (for ~1 mAh-$g_{oxide}^{-1}$) before decreasing gradually until the capacity is ~30 mAh-$g_{oxide}^{-1}$. A pseudo-plateau is observed for the following ~15 mAh-$g_{oxide}^{-1}$, following which the stress decreases consistently until the cell voltage reaches 2 V. A similar behavior is seen for the subsequent discharge cycles as well; see data for discharge cycles 2 to 4 in Figure 5. The peak compressive stress in the first cycle is ~ -6 MPa and it changes only by small amounts in the subsequent cycles. On the other hand, the stress profile during charge cycles 2 to 4 is very different from that of cycle 1. This could be attributed to the irreversible changes that occur in layered lithium metal oxide cathodes during first cycle charging which is consistent with the *in situ* XRD observations of Kang *et al.* [23]. The qualitative behavior during these cycles (2 – 4) is similar although there are quantitative differences; in all cases, the stress increases initially, then decreases later (Fig. 5b). Thus, stress in the cathode coating during the electrochemical cycling is primarily compressive except for the relatively small tensile stress during the charging process. As noted above, the measured stress represents the average stress in the composite coating, and not the stress in individual oxide particles. The stress may vary considerably within the electrode coating based on local variations in microstructure and composition. Nevertheless, the systematic changes in the average stress reveal the complex interactions between the various electrode coatings resulting from structure and volume changes not only within the oxide particles, but also within the graphite particles. It should be noted that the uncertainty in stress data presented here (Eq.2) due to estimated modulus values (according rule of mixtures, Table 2), is negligible, as a 50% change in modulus value of composite layer results in a change of less than %10 in the stress values.

The potential and stress evolution within the graphite anode coating, during the first two cycles at C/20 rate is presented in Fig. 6a and 6b, respectively. The lithiation capacity during the first cycle is ~380 mAh-g$^{-1}$ and the de-lithiation capacity is ~326 mAh-g$^{-1}$; the difference reflects the capacity associated with lithium ions trapped in the solid electrolyte interphase (SEI) layer and within the graphite particles. The potential curves show the plateau regions, both during intercalation and de-intercalation, indicating the staging behavior of graphite anode (i.e., a phase change behavior). During the lithiation process the anode coating is subjected to compressive stress which increases with capacity and reaches a peak value of – 10 MPa at the end of lithiation. Although not visible clearly, the stress behavior also showed plateau regions corresponding to staging (indicated with arrows), and is consistent with the observations of [13]. Stress evolution during delithiation, however, differs from that of the graphite anode coating in [13]. Note from Fig. 6b that the stress becomes tensile, increases with decreasing capacity, reaches a value of 4 MPa, and becomes almost constant as the delithiation process continues. This value of 4 MPa tensile stress is higher than that observed in [13] where the anode coating was mainly subjected to compressive stress during both charge and discharge processes. The reason for the quantitative differences between the present graphite anode coating and the one used in [13] may be attributed to differences in electrode constitution and porosity. For example, the anode coating in [13] had slightly higher binder content and porosity levels. However, the relative contribution of such microstructural parameters to the observed difference in the electrode stress evolution is unknown. This observation, nevertheless, suggests that processing and microstructural parameters such as porosity level and the proportions of individual constituents are very important in determining the stress and the consequent mechanical degradation of composite electrodes.

**3.2 Pressure exerted by spirally wound electrodes on battery casing**

Stresses in electrode coatings result in internal pressure on the battery casing, particularly in the cylindrical spirally wound configuration (i.e. the jelly roll). There are multiple phenomena that contribute to internal pressure on a battery casing during cycling, such as gas evolution, temperature changes, etc. The objective here is to investigate the contribution of stress in the jelly-roll electrode coating to the internal pressure on the casing, i.e, the compressive stress induced between the outer most layer of the jelly-roll and the casing. Fig. 7 shows the stress evolution in cathode and anode coating as a function of cell capacity in the first and second cycles, which is used in the following analysis to estimate the internal pressure that would result in a spirally wound battery casing. The classical solution for thermal stresses in a thick walled cylinder [24] is extended here to multiple concentric cylinders to calculate the pressure exerted by electrodes on the battery casing. The expressions for radial displacement $u$ and radial stress $\sigma_r$ in a long cylinder (plane strain approximation) due to an axisymmetric temperature distribution are given by,

$$u = \frac{1+v}{1-v}\alpha \frac{1}{r}\int_{r_i}^{r} T(r)r\, dr + C_1 r + \frac{C_2}{r}, \quad\quad\quad\quad\quad\text{Eq.3}$$

$$\sigma_r = -\frac{\alpha E}{1-v}\frac{1}{r^2}\int_{r_i}^{r} T(r)r\, dr + \frac{E}{1+v}\left(\frac{C_1}{1-2v} - \frac{C_2}{r^2}\right), \quad\quad\text{Eq.4}$$

where $E$, $v$, and $\alpha$ are Young's modulus, Poisson's ratio, and coefficient of thermal expansion, respectively. $T(r)$ is the radial temperature distribution; $C_1$ and $C_2$ are constants determined by the boundary conditions at the inner and outer surfaces of the cylinder; $r_i$ is the inner radius of the cylinder. Exploiting the similarity between thermal stresses and lithiation induced stresses [25], the above equations can be written as,

$$u = \frac{1+v}{1-v}\frac{\varepsilon^*}{r}\int_{r_i}^{r} r\, dr + C_1 r + \frac{C_2}{r}, \quad\quad\quad\quad\text{Eq.5}$$

$$\sigma_r = -\frac{E}{1-v}\frac{\varepsilon^*}{r^2}\int_{r_i}^{r} r\, dr + \frac{E}{1+v}\left(\frac{C_1}{1-2v} - \frac{C_2}{r^2}\right), \quad\quad\text{Eq.6}$$

where $\varepsilon^*$ is the eigen strain due to the uniform lithium concentration in the electrode (i.e. the strain due to lithiation), which takes the place of the thermal strain $\alpha T(r)$; it is assumed that the lithium concentration is uniform throughout the thickness of the cylinder, which represents a single electrode layer. The eigen strain in cathode (or anode) coatings at any given capacity can be obtained as,

$$\varepsilon^* = \sigma(1-v)/E, \quad\quad\quad\text{Eq.7}$$

where $\sigma$ is the measured stress in the electrode coatings (e.g. Fig. 7). $E$ and $v$ are the Young's modulus and the Poisson's ratio of the coating. The Eigen strain for the current collector layers (copper and aluminum) and the separator is taken as zero since they do not participate in the electrochemical reactions. These equations (Eqs. 5, 6, and 7) assume that the mechanical properties ($E$ and $v$) of the coatings are independent of the state of charge and are isotropic; also, the Eq. 7 is valid for linear elastic material behavior. It should be noted that the precise values of the Young's modulus ($E$) and Poisson's ratio ($v$) for the cathode coating are not available. However, several studies indicated that the $E$ and $v$ values of layered oxide materials such as $LiCoO_2$ are generally in the range of 85-180 GPa and 0.3, respectively [11]; hence, a representative value of 100 GPa for the secondary cathode particles was assumed. The property values for the cathode (and anode) coating are obtained by assuming a rule of mixtures with the cathode (and anode) particles, PVdF binder and the known porosity as the constituents.

Fig. 8 schematically represents the cross section of a spirally wound battery in which the spirals are approximated by concentric circles, which is reasonable as the thickness of any individual layer is relatively small (<35 µm) compared to the overall radius of the battery (*e.g.,* 9.3 mm for the 18650 battery); note that an 8-layer stack consisting of an Al current collector with cathode coating on both sides, a Cu current collector with anode coating on both sides and 2 separator layers – forms a repeating unit in the radial direction. Eqs. 5 and 6 can be extended to each layer of Fig. 8 with the following boundary conditions, $\sigma_r = 0$ at $r = r_i$, the inner most radius of the spiral; $u = 0$ at $r = r_o$, the outer radius, which is also the inner radius of the battery casing (i.e., the battery casing is assumed to be rigid); and

$$\sigma_r^L = \sigma_r^{L+1}$$

$$u^L = u^{L+1}$$

at the interface between layers $L$ and $L+1$ for L= 1,...., $n-1$ where $n$ is the number of individual layers (i.e., there are in total 304 layers in our model which considers an 18650 battery casing and the coating thickness of the anode and cathode on which stress measurements are reported above). The resulting system of linear equations is solved for the constants $C_1$ and $C_2$ for each layer at a given cell capacity, which can then be used to determine the radial stress and displacement at all radial locations. The pressure exerted by the electrode winding at a given capacity on the battery casing is obtained by calculating $\sigma_r$ at $r = r_o$. As mentioned earlier, the material properties of each individual layer (given in Table 1 and 2) are assumed to be independent of cell capacity. Hence, the evolution of the pressure (i.e., $\sigma_r$ at $r = r_o$) depends only on the Eigen strain in the electrodes (see Eq.6), which is obtained directly from the experimental measurement shown in Fig. 7.

Fig. 9 shows the calculated evolution of internal pressure on the battery casing as a function of cell capacity for two charge/discharge cycles. As expected, the pressure on the casing varies continuously with the capacity. The pressure exerted by electrodes due to electrolyte soaking alone is 0.15 MPa, and it increases to a peak value of approximately 1 MPa at the end of charging. Upon discharge the pressure decreases rapidly as seen in Fig. 9, relieving the internal pressure. Although the model predicts a slightly negative pressure during discharge process, it is an artifact of the simple model used here in which perfect bonding between the casing and outer most electrode layer is implied by the boundary conditions employed, which is applicable only when the electrode presses against the casing (i.e., when the pressure is positive). Since the main objective of calculating the pressure is to estimate the peak positive pressure and the resulting peak stress in the casing, the calculated negative pressure should be interpreted as zero pressure. The pressure during the second charge process is lower than the first cycle pressure. It can be safely assumed that the peak pressure in the subsequent cycles will be smaller, unless the stress in either cathode coating or anode coating increases in magnitude during subsequent cycles. The circumferential (or hoop) stress in the casing is given by,

$$\sigma_\theta = \frac{pr_o}{t}, \quad \ldots\ldots\ldots\ldots\ldots\ldots\ldots\ldots\ldots\ldots\ldots \text{Eq. 8}$$

by assuming the casing as a thin walled cylinder. Here, $p$ is internal pressure; $r_o$ and $t$ are radius and thickness of battery casing (Fig.8), respectively. For a casing with 9.3 mm radius and 100 μm assumed thickness, an internal pressure of ~1 MPa results in 93 MPa of hoop stress according to Eq.8. As noted earlier, the pressure of ~1 MPa and the calculated hoop stress value are only lower bound estimates as there could be other sources of pressure such as gas evolution [26, 27] which are not considered in this study. Using typical strength values of 200 MPa and 500 MPa for aluminum and steel respectively, the hoop stress estimate above gives a factor of safety of 2.2 and 5.4 for the two materials. It must be noted that the assumptions employed in the model, i.e., isotropic electrode swelling, uniform lithiation throughout the electrodes, Eigen strain from Eq.7 which is valid only for linear elastic behavior, and approximate material properties of Table 2 render the pressure values calculated to be first order estimates.

Nevertheless, these values provide valuable information for the design of battery electrodes, packs and casings.

## 4 Conclusions

Real-time average stress evolution in composite cathode coatings made of $Li_{1.2}Ni_{0.15}Mn_{0.55}Co_{0.1}O_2$ secondary particles and anode coatings made of graphite particles were measured by substrate curvature technique. The cathode was paired with a capacity-matched composite graphite anode to mimic the conditions that the cathode material is subjected to in a practical application. The stress evolution due to solvent absorption was monitored until it reached a steady-state (approximately 60 min) prior to any electrochemical cycling. The stress evolution is observed to be approximately linear at the beginning and reached a plateau value of about -0.15 MPa after 60 min of soaking.

Followed by electrolyte soaking, the cells were charged/discharged galvanostatically at a current density of 0.1 mA/cm$^2$, which corresponds to a rate of C/20. The first cycle charge capacity of the cathode was 311 mAh/g, which is comparable to that obtained from a coin-cell experiment. It was observed that the stress evolution during the initial charge/discharge cycle is not monotonic suggesting complex changes in the volume of the active material. The stress and potential evolution during the initial charge cycle is different from that of the other charge cycles which is consistent with XRD studies on similar layered cathodes. The qualitative stress behavior is repeated in the subsequent cycles (*i.e.*, 2, 3, and 4$^{th}$ cycle), although there are some quantitative differences. The stress is primarily compressive except for the relatively small tensile stress during the charging process. The peak compressive stress in the first cycle is approximately -6 MPa and it changes only by a small amount in the subsequent cycles. The changes in the average stress possibly suggest complex crystal structure and volume changes in the active material particles; these changes in stress evolution data suggest that a more comprehensive XRD study on individual primary particles should be undertaken to better understand the response of $Li_{1.2}Ni_{0.15}Mn_{0.55}Co_{0.1}O_2$ electrodes. The stress evolution in the graphite anode coating prepared here differed slightly from the previously reported behavior in [13] which can be attributed to the different processing parameters of the coatings. This observation suggests that the stress evolution and associated phenomena are dependent on the microstructure and process parameters of the coatings.

For the cathode and anode coatings considered above, a model is presented to calculate the internal pressure that would be generated in a spirally wound battery casing due to the electrode stress during charge/discharge cycling. Using the experimentally measured stress values, the internal pressure in an 18650 battery casing is calculated. As expected, the pressure on the casing varies with the cell capacity. The pressure exerted by electrodes due to electrolyte soaking is ~0.15 MPa and increases to a peak value of approximately 1 MPa at the end of the first charge. The pressure is relieved during most of the discharge. The peak pressure during subsequent cycles is expected to remain lower than that of the first cycle. Hence, the pressure estimated during the first cycle is sufficient to consider the contribution of electrode stress on the hoop stress in the battery casing. The experiments and calculations presented here are expected to provide useful information for the design of electrode coatings, packs and casings.

## 5. Acknowledgements

The authors gratefully acknowledge financial support from the United States Department of Energy – EPSCoR Implementation award (grant # DE-SC0007074). We are also grateful to the U.S. Department of Energy's (DOE) Cell Analysis, Modeling and Prototyping (CAMP) Facility at Argonne National Laboratory for providing electrodes used for this work. The CAMP facility is fully supported by DOE's Vehicle Technologies Program (VTP) within the core funding of the Applied Battery Research (ABR) for Transportation Program. SN acknowledges the financial support provided by New Jersey Institute of Technology through the faculty startup research grant.

**Figures**

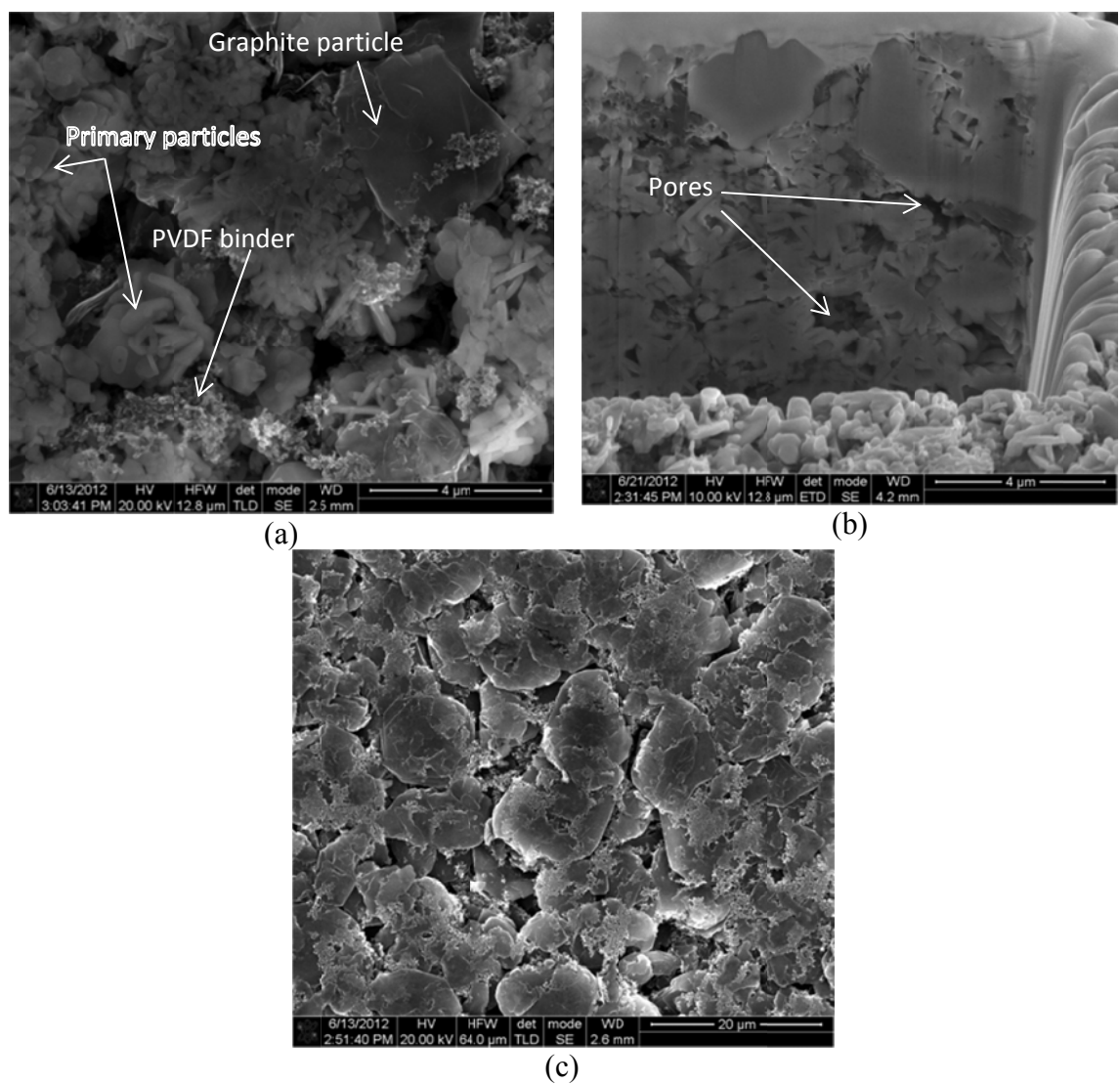

Fig. 1: As-prepared microstructure of $Li_{1.2}Ni_{0.15}Mn_{0.55}Co_{0.1}O_2$ cathode coating showing (a) secondary particles, primary particles, and SFG-6 graphite particle. (b) A cross-sectional view of the cathode revealing substantial porosity (c) As-prepared microstructure of graphite anode coating.

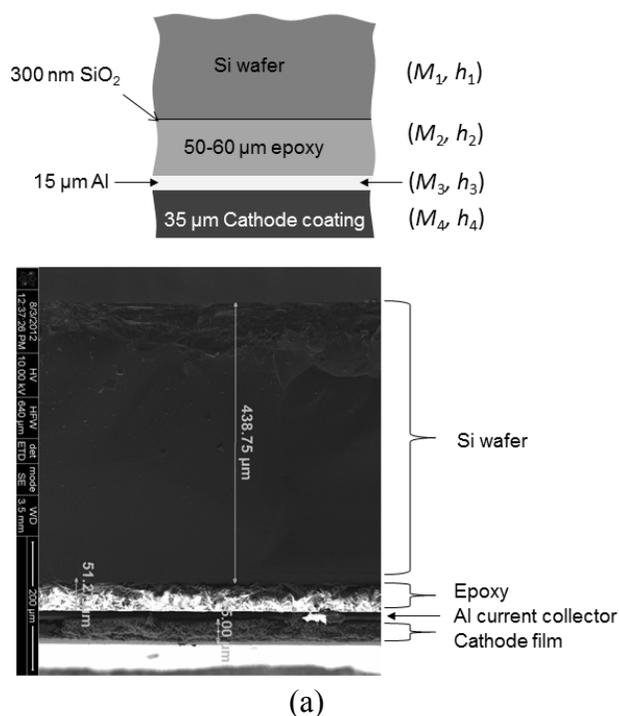

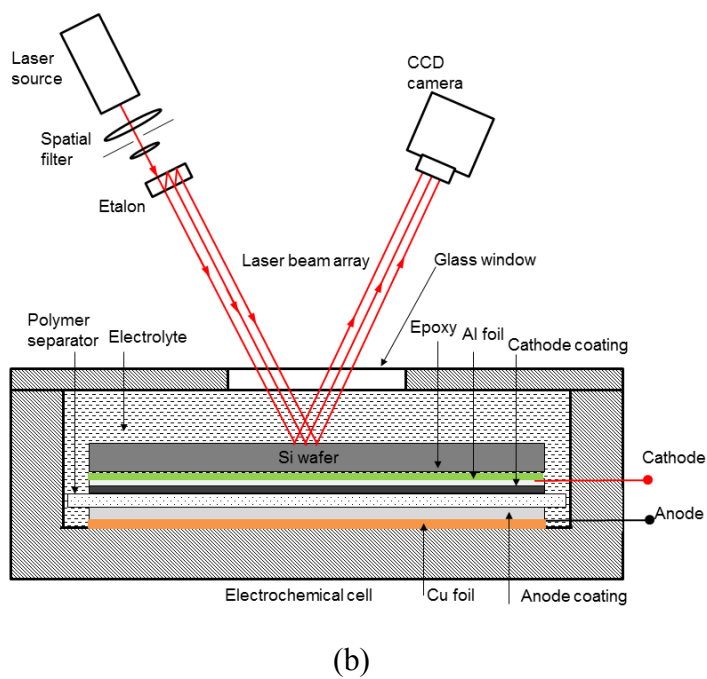

Fig.2: (a) Schematic of the sample cross-section and the corresponding SEM image showing the substrate, epoxy, Al current collector, and cathode coating. The thickness of each layer is also shown. The thicknesses of the layers are $h_1$-$h_4$ and the corresponding bi-axial moduli are $M_1$-$M_4$. The bi-axial modulus is defined as M= E/(1-v) where E is the Young's Modulus and v is Poisson's ratio of the material. (b) Schematic of experimental setup showing the multi-beam optical sensor for substrate curvature measurements integrated with the electrochemical cell.

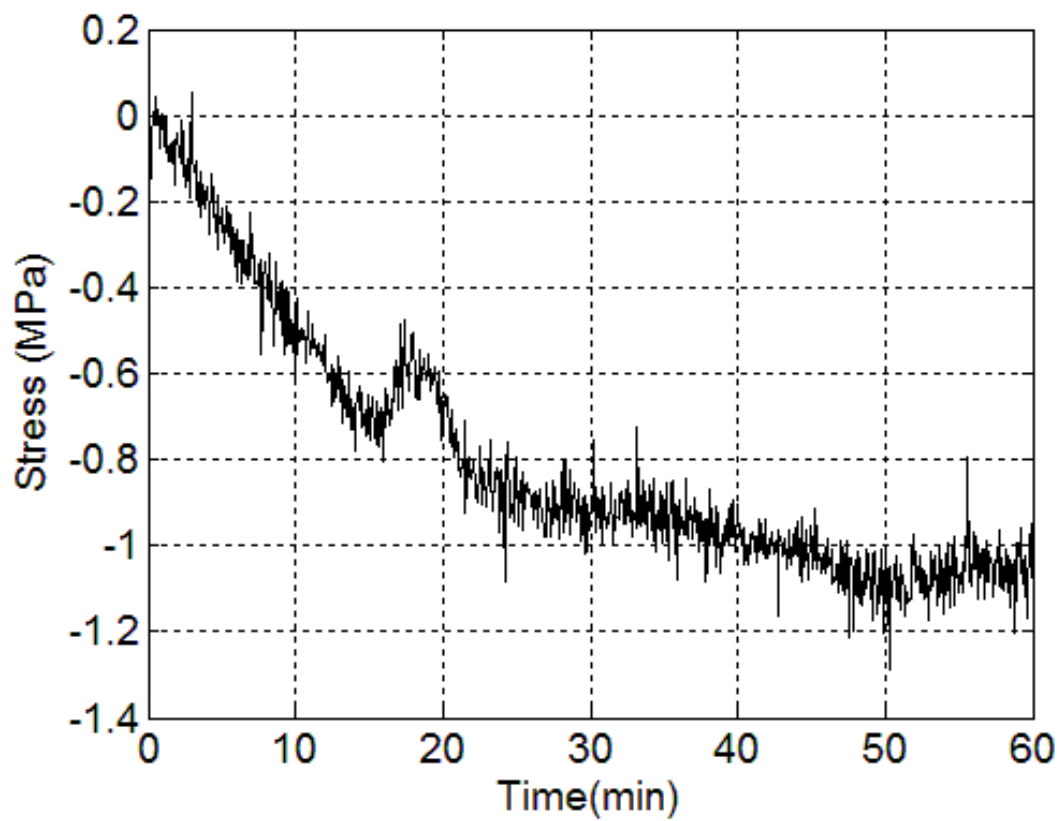

Fig. 3: Typical stress evolution during the initial intake of electrolyte into the porous regions of cathode and swelling of binder.

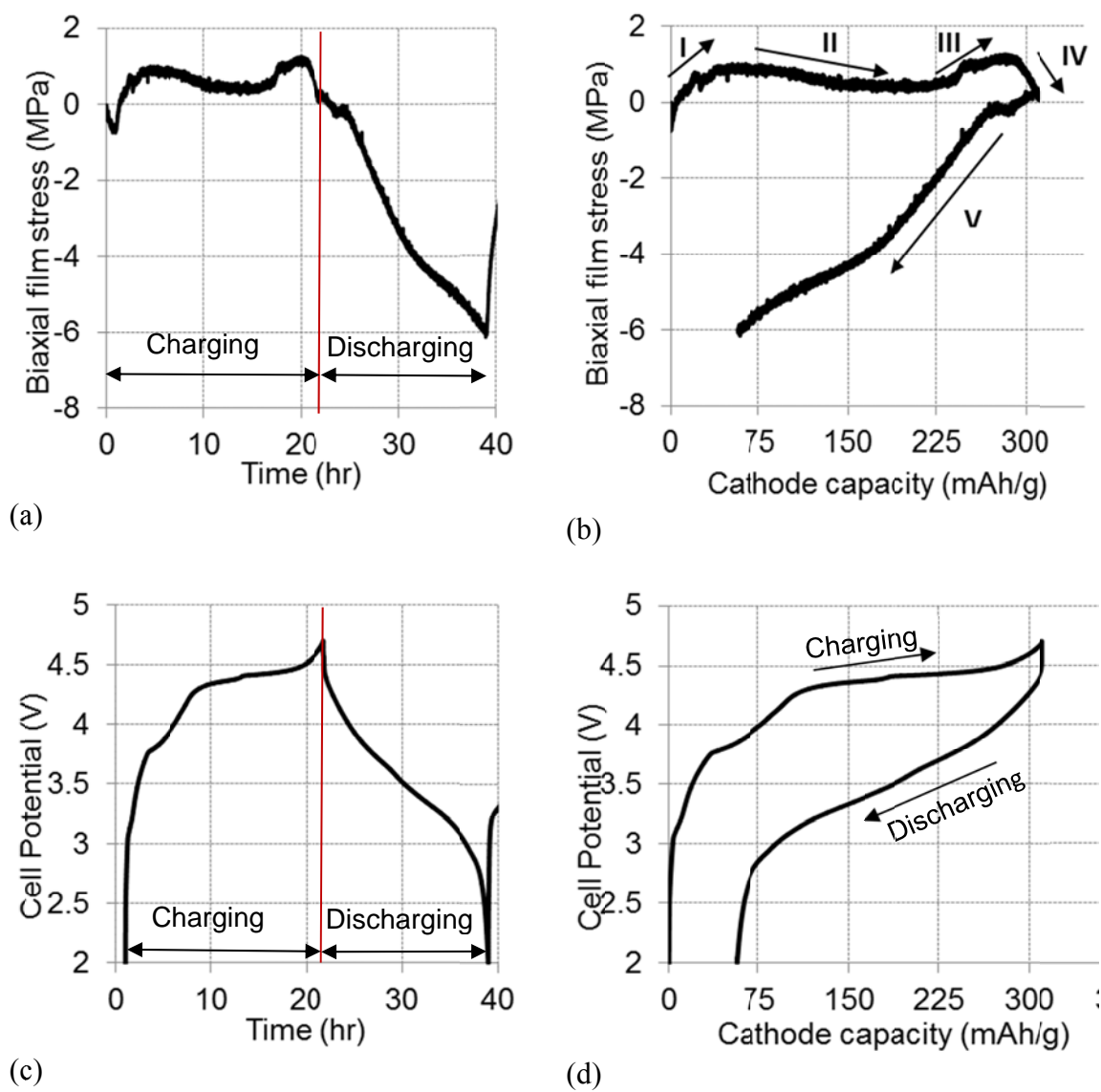

Fig. 4: (a)-(b) show the "average" stress evolution in cathode coating, and (c)-(d) show potential evolution of full cell during the first charge-discharge (C/20 rate) cycle as a function of time and capacity, respectively. (b) Shows different regimes of changes (*i.e.*, increase and decrease) in the cathode stresses which could be due to volume changes associated with phase transformation. Li removal from/insertion into cathode coating is defined as charging/discharging respectively.

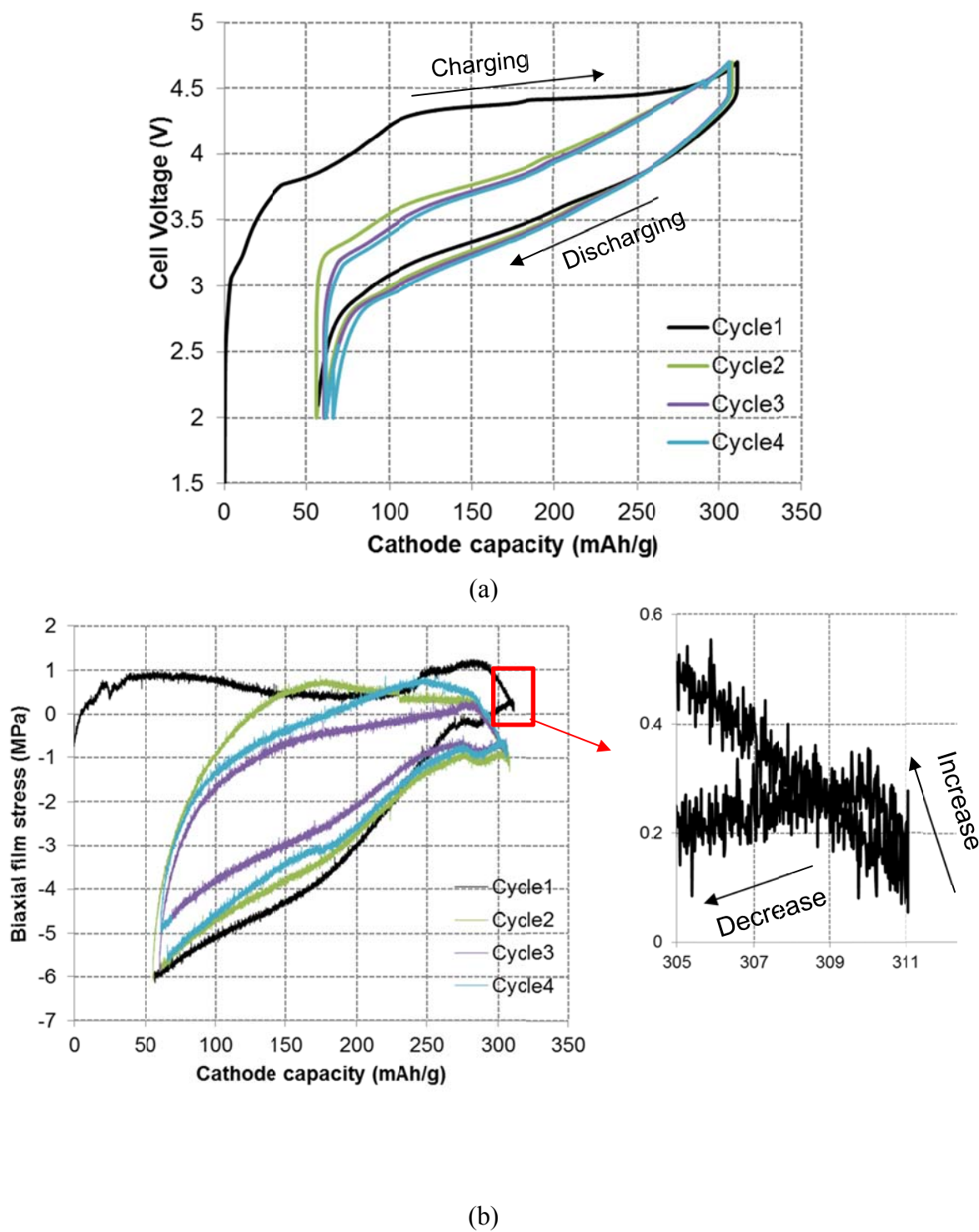

Fig. 5: (a) Potential evolution of full cell, and (b) the stress evolution of cathode as a function of capacity (full cell, C/20 rate). The inset shows that the stress starts to increase before decreasing at the beginning of delithiation.

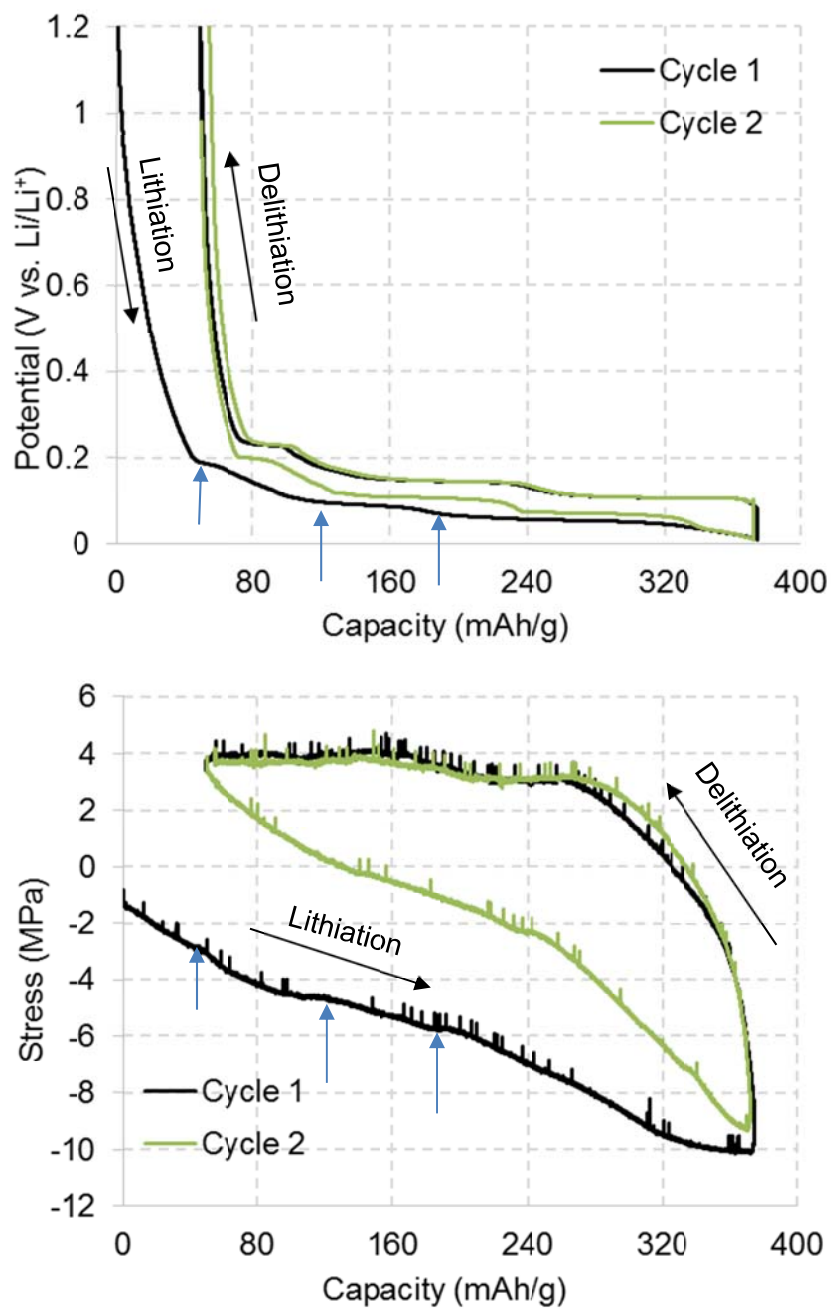

Fig. 6: Potential and average stress evolution of anode coating as a function of capacity during the first two cycles. The staging behavior of graphite is reflected in stress evolution as well.

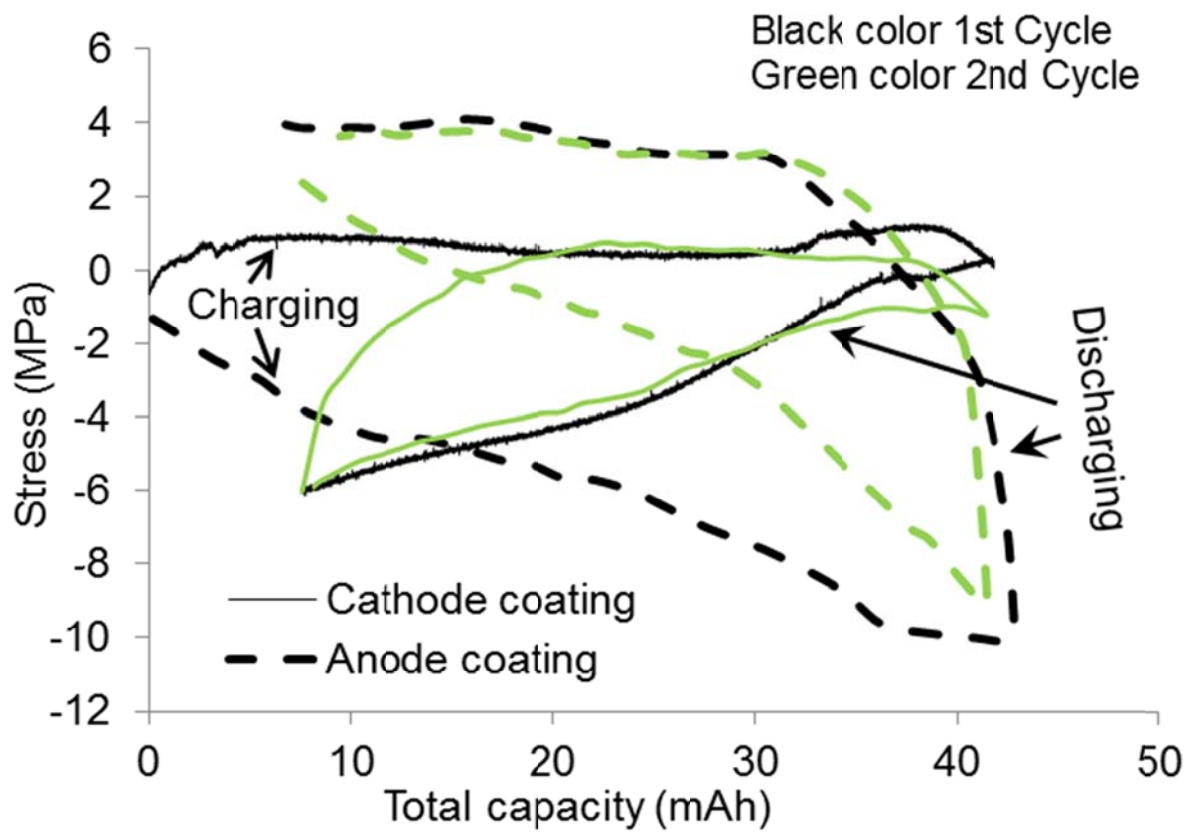

Fig. 7: Stress evolution in cathode and anode coatings as a function of cell capacity during the first (black curves) and second (green curves) charge/discharge cycles. Solid lines and dashed lines represent stress in cathode and anode coatings respectively.

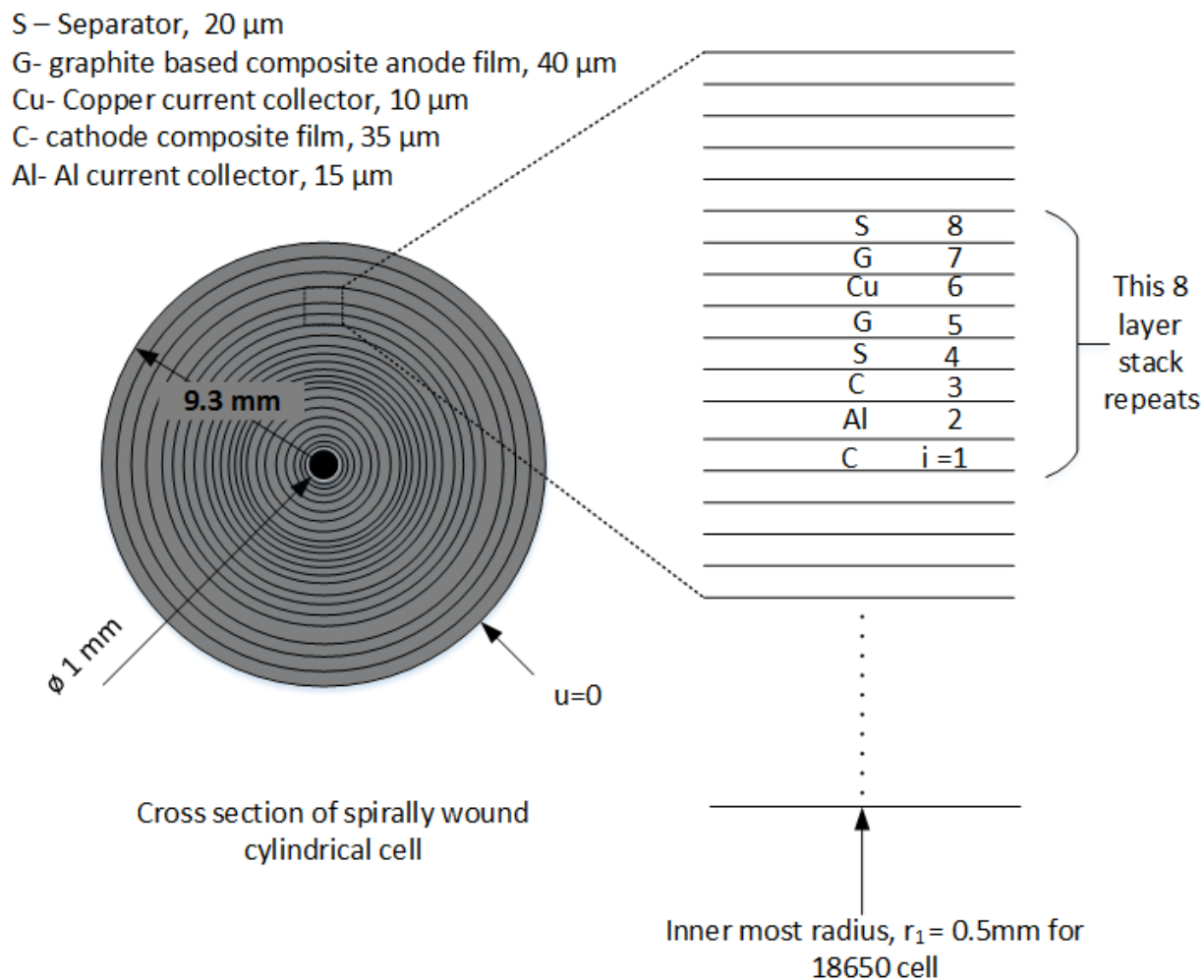

Fig. 8: Schematic showing the cross section of spirally wound 18650 lithium-ion battery along with geometric details of different layers. For the purposes of this analysis, the spirals are approximated as concentric circles. The inner and outer radius of the battery are $r_i = 0.5$ and $r_o = 9.3$ mm, respectively. For the thicknesses of individual layers considered in this investigation (Tables 1 and 2), our model jelly-roll battery has 304 layers.

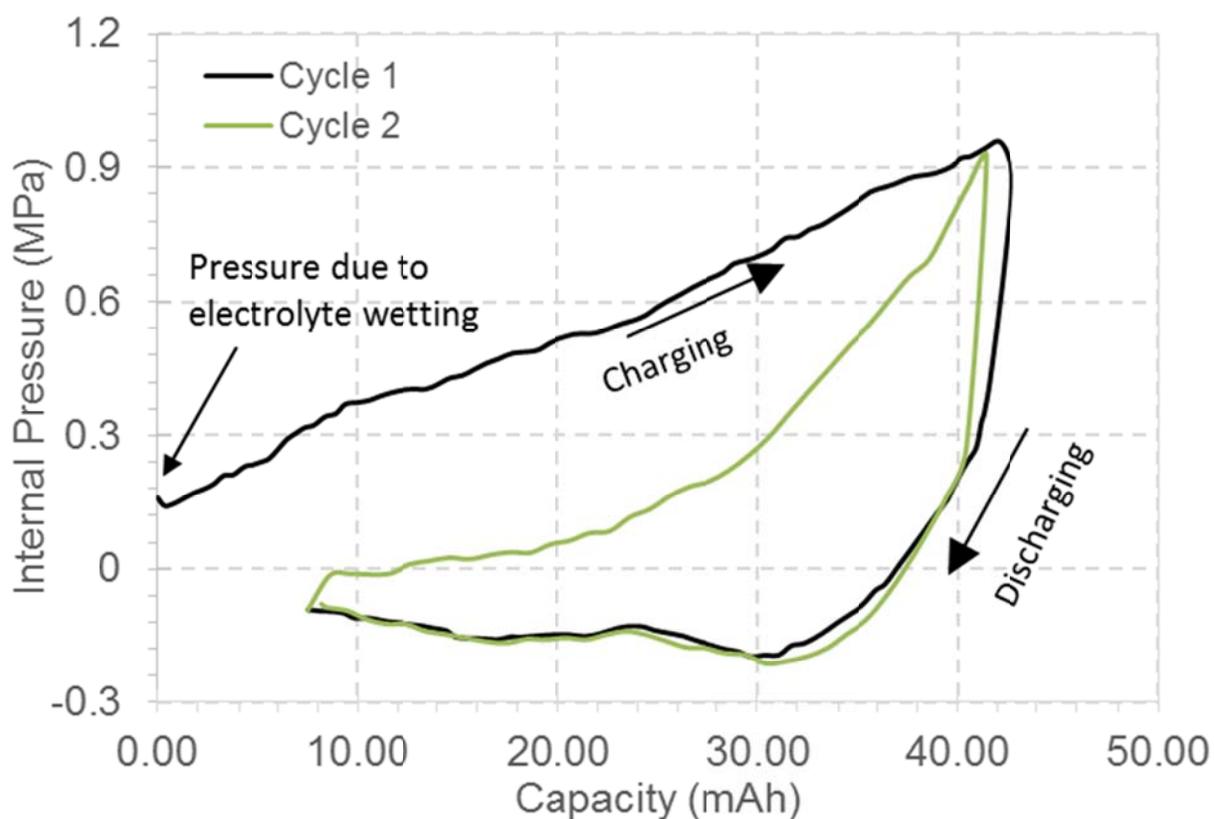

Fig. 9: Calculated variation of internal pressure as a function of electrode capacity for a spirally wound battery made of the cathode and anode coatings considered above in a 18650 cell configuration. The pressure increases almost linearly with capacity during the first charge and it is relieved rapidly during the discharge. The negative pressures should be interpreted as zero pressure. The peak pressure is ~1 MPa.

## Tables

**Table 1: Properties of composite electrodes and the proportions of their individual constituents [20]**

| Description | Value |
|---|---|
| ***I.     Cathode*** | |
| Toda-HE-5050: $Li_{1.2}Ni_{0.15}Mn_{0.55}Co_{0.1}O_2$ | 86% wt. |
| Solvay 5130 PVDF binder | 8% wt. |
| Timcal SFG-6 graphite | 4% wt. |
| Timcal Super P | 2 % wt. |
| Active-material loading density | 6.64 mg/cm$^2$ |
| Electrode porosity | 37.1% |
| Thickness of the coating | 35 μm |
| Thickness of the Al current collector | 15 μm |
| ***II.    Anode*** | |
| ConocoPhillips: CGP-A12 graphite | 89.8% wt. |
| KF-9300 Kureha PVDF binder | 6% wt. |
| Timcal Super P | 4%wt. |
| Oxalic Acid | 0.17% wt. |
| Active-material loading density | 5.61 mg/cm$^2$ |
| Electrode porosity | 26% |
| Thickness of the coating | 40 μm |
| Thickness of the Cu current collector | 10 μm |

**Table 2: Mechanical properties and geometry of different layers of the specimen used in the study**

| Parameter | Definition | Value | Comments |
|---|---|---|---|
| **Si (111) wafer** | | | |
| $E_1$ | Young's modulus | 169 GPa | |
| $v_1$ | Poisson's ratio | 0.26 | |
| $h_1$ | Thickness | 450 μm | Measured |
| $M_1$ | Biaxial modulus | 228.3 GPa | Calculated |
| **Epoxy layer** | | | |
| $E_2$ | Young's modulus | 4.3 GPa | Ref.[13] |
| $v_2$ | Poisson's ratio | 0.36 | |
| $h_2$ | Thickness | 55 μm | Measured |
| $M_2$ | Biaxial modulus | 6.72 GPa | Calculated |
| **Al current collector** | | | |
| $E_3$ | Young's modulus | 70 GPa | |
| $v_3$ | Poisson's ratio | 0.334 | |
| $h_3$ | Thickness | 15 μm | Measured |
| $M_3$ | Biaxial modulus | 105 GPa | Calculated |
| **Composite cathode coating** | | | |
| $E_4$ | Young's modulus | 40 GPa | Rule of mixtures |
| $v_4$ | Poisson's ratio | 0.2 | Rule of mixtures |
| $h_4$ | Thickness | 35 μm | Measured |
| $M_4$ | Biaxial modulus | 50 GPa | Calculated |
| **Other parameters** | | | |
| $\cos\phi/2L$ | Mirror constant | 2 | Measured |
| $d_f$ | Diameter of specimen | 50.8 mm | Measured |
| **Graphite based anode coating** | | | |
| $E_1$ | Young's modulus | 6.9 GPa | Ref. [13] |
| $v_1$ | Poisson's ratio | 0.3 | Ref. [13] |
| $h_1$ | Thickness | 35 μm | |
| $M_1$ | Biaxial modulus | 10 GPa | Calculated |
| **Copper current collector** | | | |
| $E_1$ | Young's modulus | 117 GPa | |
| $v_1$ | Poisson's ratio | 0.347 | |
| $h_1$ | Thickness | 15 μm | |
| $M_1$ | Biaxial modulus | 179 GPa | Calculated |
| **Celgard separator** | | | |
| $E_1$ | Young's modulus | 0.1 GPa | Ref. [28] |
| $v_1$ | Poisson's ratio | 0.3 | |
| $h_1$ | Thickness | 20 μm | Measured |
| $M_1$ | Biaxial modulus | 0.14 GPa | Calculated |

**Figure captions**

Fig. 1: As-prepared microstructure of $Li_{1.2}Ni_{0.15}Mn_{0.55}Co_{0.1}O_2$ cathode coating showing (a) secondary particles, primary particles, and SFG-6 graphite particle. (b) A cross-sectional view of the cathode revealing substantial porosity (c) As-prepared microstructure of graphite anode coating.

Fig.2: (a) Schematic of the sample cross-section and the corresponding SEM image showing the substrate, epoxy, Al current collector, and cathode coating. The thickness of each layer is also shown. The thicknesses of the layers are $h_1$-$h_4$ and the corresponding bi-axial moduli are $M_1$-$M_4$. The bi-axial modulus is defined as M= E/(1-v) where E is the Young's Modulus and v is Poisson's ratio of the material. (b) Schematic of experimental setup showing the multi-beam optical sensor for substrate curvature measurements integrated with the electrochemical cell.

Fig. 3: Typical stress evolution during the initial intake of electrolyte into the porous regions of cathode and swelling of binder.

Fig. 4: (a)-(b) show the "average" stress evolution in cathode coating, and (c)-(d) show potential evolution of full cell during the first charge-discharge (C/20 rate) cycle as a function of time and capacity, respectively. (b) Shows different regimes of changes (*i.e.*, increase and decrease) in the cathode stresses which could be due to volume changes associated with phase transformation. Li removal from/insertion into cathode coating is defined as charging/discharging respectively.

Fig. 5: (a) Potential evolution of full cell, and (b) the stress evolution of cathode as a function of capacity (full cell, C/20 rate). The inset shows that the stress starts to increase before decreasing at the beginning of delithiation.

Fig. 6: Potential and average stress evolution of anode coating as a function of capacity during the first two cycles. The staging behavior of graphite is reflected in stress evolution as well.

Fig. 7: Stress evolution in cathode and anode coatings as a function of cell capacity during the first (black curves) and second (green curves) charge/discharge cycles. Solid lines and dashed lines represent stress in cathode and anode coatings respectively.

Fig. 8: Schematic showing the cross section of spirally wound 18650 lithium-ion battery along with geometric details of different layers. For the purposes of this analysis, the spirals are approximated as concentric circles. The inner and outer radius of the battery are $r_i$ = 0.5 and $r_o$ = 9.3 mm, respectively. For the thicknesses of individual layers considered in this investigation (Tables 1 and 2), our model jelly-roll battery has 304 layers.

Fig. 9: Calculated variation of internal pressure as a function of electrode capacity for a spirally wound battery made of the cathode and anode coatings considered above in a 18650 cell configuration. The pressure increases almost linearly with capacity during the first charge and it is relieved rapidly during the discharge. The negative pressures should be interpreted as zero pressure. The peak pressure is ~1 MPa.